\documentclass[12pt]{iopart} 
\usepackage{iopams}  
\usepackage{epsf}
\begin{document} 
\jl{1} 
 
\paper[Non-linear coherent states]{Non-linear coherent states associated with
  conditionally exactly solvable problems} 
\author{Georg Junker\dag\footnote[3]{E-mail:
    junker@theoriel.physik.uni-erlangen.de} and Pinaki
  Roy\ddag\footnote[4]{E-mail: pinaki@isical.ac.in}}  
 
\address{\dag\ Institut f\"ur Theoretische Physik, 
Universit\"at Erlangen-N\"urnberg, Staudtstr.\ 7, D-91058 Erlangen, Germany} 
 
\address{\ddag\ Physics and Applied Mathematics Unit,
Indian Statistical Institute, Calcutta 700035, India} 
 
\begin{abstract} 
Recently, based on a supersymmetric approach, new classes of conditionally
exactly solvable problems have been found, which exhibit a symmetry structure
characterized by non-linear algebras. In this paper the associated
``non-linear'' coherent states are constructed and some of their properties
are discussed in detail.
\end{abstract} 
~

\section{Introduction}
It is well known that only a few quantum mechanical models admit exact
solutions. The class of exactly solvable models can, however, be enlarged by
using the technique of generating isospectral Hamiltonians \cite{1}. Recently,
another class of problems \cite{2,3} consisting of so-called conditionally
exactly 
solvable (CES) problems has emerged. The characteristic feature of this class
is that their members are exactly solvable problems when the parameters
appearing in the potential are fine tuned to assume some specific numerical
value or to lie in some range of values. 

In some recent papers \cite{4,5,5a} several classes of CES problems, whose
construction is based on supersymmetric (SUSY) quantum mechanics \cite{7} have
been 
found. It was shown \cite{4,5} that the classes associated with the linear and
radial harmonic oscillator admit some non-linear algebra as their symmetry
algebra. Here our objective is to construct coherent states corresponding to
these CES problems and examine some of their properties. We recall that 
usually coherent states are constructed using as a basis some Lie algebra
\cite{6}. In contrast, here the coherent states are constructed over
non-linear algebras and we call them {\em non-linear coherent states}. In this
paper we limit ourselves to the class of CES potentials associated with the
radial harmonic oscillator. 
To be more precise, we shall start with systems having $su(1,1)$
dynamical symmetry and then construct coherent states corresponding to
the isospectral partners which have a non-linear (i.e.\ deformed)
$su(1,1)$ symmetry. In this context we recall that in ref \cite{Nie79}
Nieto et al described a method of constructing coherent and squeezed
states for arbitrary quantum mechanical potentials. In the present
paper we construct coherent states for hitherto unknown potentials
having some particular symmetry properties. To be a bit more explicit, we
consider two cases. One in which SUSY is broken and the other in which SUSY is
unbroken. In the former case non-linear coherent states can be constructed
over the entire Fock space. Whereas in the latter case non-linear coherent
states are defined in a subspace of the Fock space. 

This paper is organized as follows. In the next section we briefly summarize
the essentials of SUSY quantum mechanics. In Section 3 we discuss the CES
potentials associated with the radial harmonic oscillator model and its
non-linear symmetry algebra. Section 4 is devoted to the construction of the
non-linear coherent states. Basic properties of these states are also
discussed. Finally, in Section 5 we briefly discuss the case of unbroken SUSY
and in Section 6 some discussion and outlook is given.
\section{SUSY quantum mechanics}
To begin with we note that Witten's model of supersymmetric quantum mechanics
consists of a pair of Hamiltonians \cite{7}
\begin{equation}
H_{\pm}=- \frac{1}{2}\, \frac{\rmd^2}{\rmd x^2} + V_{\pm} (x)
\end{equation}
acting on some suitable Hilbert space ${\cal H}$. For the purpose at hand we
take 
the linear space of square integrable functions on the positive half-line with
Dirichlet boundary condition at the origin, 
${\cal H} = \left\{ \psi \in L^2 ({\mathbb R}^{+}) | \psi (0) = 0 \right\}$.
The supersymmetric partner potentials in (1) are given by
\begin{equation}
V_{\pm} (x) = \frac{1}{2}\left[W^2 (x) \pm W'(x)\right]
\end{equation}
where $W$ is the SUSY potential and $W'=dW/dx$. In terms of the operators
\begin{equation}
A = \frac{1}{\sqrt{2}}\, \left( \frac{\rmd}{\rmd x} + W(x) \right) \;,
\quad A^{\dagger} = \frac{1}{\sqrt{2}}\, \left( - \frac{\rmd}{\rmd x} + W(x)
\right)  
\end{equation}
the Hamiltonians in (1) read $H_{+} = A A^{\dagger}$ and $H_{-} =  A^{\dagger}
A$, respectively. Let us denote the eigenfunctions and eigenvalues of
$H_{\pm}$ by  $\psi^{\pm}_{n}$ and $E^{\pm}_{n}$:
\begin{equation}
H_{\pm} \psi^{\pm}_{n} (x) = E^{\pm}_{n} \psi^{\pm}_{n} (x)\;,\quad n = 0,
1, 2, \ldots\;.
\end{equation}
Then it can be shown \cite{7} that in the case of broken SUSY (we will mainly
concentrate on this case) the spectrum of $H_{-}$ coincides with that of
$H_{+}$ and both are strictly positive:
\begin{equation}
\fl E^{+}_{n} = E^{-}_{n} \equiv E_n > 0\;, \quad
\psi^{-}_{n} (x) = E_{n}^{-1/2} A^{\dagger} \psi^{+}_{n}(x)\;,
\quad \psi^{+}_{n} (x) = E_{n}^{-1/2} A \psi^{-}_{n} (x)\;.
\end{equation}
Thus it is clear that if one of the Hamiltonians is exactly solvable
then the spectral properties of the other one are also known, that is, it
becomes exactly solvable, too. This is the basic idea in the supersymmetric
construction methods of CES potentials. To be a bit more
explicit, in \cite{4,5,5a} it has been suggested to construct SUSY potentials
$W$ in such a 
way that $V_+$ becomes (under certain conditions imposed on the parameters
involved) one of the well-known exactly solvable (shape-invariant) potentials
and thus giving rise, in general, to a class of CES potentials $V_-$.

\section{A model with broken SUSY}
Now as a specific model we consider the following SUSY potential
\begin{equation}
W(x) = x + \frac{\gamma + 1}{x} + \frac{u'(x)}{u(x)}\;,
\end{equation}
where
$u(x)={}_1F_1(\textstyle\frac{1-\varepsilon}{2},\gamma+\frac{3}{2},-x^2)$ is a
confluent hypergeometric function and
the two potential parameters have to obey the conditions $\gamma\geq 0$ and
$\varepsilon > -2\gamma -2$.
This SUSY potential can be shown \cite{5,5a} to give rise to 
\begin{equation}
V_{+}(x) = \frac{x^2}{2} + \frac{\gamma(\gamma + 1)}{2x^2} + \varepsilon +
\gamma + \frac{1}{2}\;.
\end{equation}
Clearly, $V_{+}$ represents the generalised radial harmonic
oscillator (this can be regarded as the potential corresponding to the two body
Calogero-Sutherland model) and the associated spectral properties of $H_+$ are
well known 
\begin{equation}
\fl E_{n} = 2n + 2 \gamma + 2 + \varepsilon\;,\quad
\psi^{+}_{n} (x) = \left[ \frac{2n !}{\Gamma(n + \gamma + \frac{3}{2})}
\right]^{1/2} x^{\gamma + 1}\,\rme^{- x^2 / 2}\,L_{n}^{\gamma + \frac{1}{2}}
(x^2)\;.
\end{equation}
Here $L_n^\nu$ denotes a generalised Laguerre polynomial and we also note
that SUSY is broken, that is, 
$\exp\left\{\pm\int\rmd x\, W(x)\right\}\notin {\cal H}$.
As a consequence, the SUSY partner Hamiltonian $H_-$ has the same eigenvalues
$E_n$ and its eigenfunctions can be obtained from (8) via (5):
\begin{equation}
\fl
\begin{array}{rl}
\psi^{-}_{n} (x) &= \displaystyle
\frac{1}{\sqrt{4n + 4\gamma + 4 + 2\varepsilon}} 
\left(-\frac{\rmd}{\rmd x} + x + \frac{\gamma+1}{x} + \frac{u'(x)}{u(x)}\right)
\psi^{+}_{n} (x)\\[4mm]
&\hspace{-8mm}=\displaystyle\left[\frac{2\,n!}
{(n+\gamma+1+\frac{\varepsilon}{2})\Gamma(n+\gamma+\frac{3}{2})}\right]^{1/2}
x^{\gamma+2}\,\rme^{-x^2/2}
\left(L_n^{\gamma+3/2}(x^2)+\frac{u'(x)}{2\,x\,u(x)}\right)\;.
\end{array}
\end{equation}
The corresponding CES potential explicitly reads
\begin{equation}
\fl V_{-} (x) = \frac{x^2}{2} + \frac{(\gamma+1) (\gamma+2)}{2x^2} + \gamma -
\varepsilon + \frac{3}{2} + \frac{u'(x)}{u(x)} \left( 2x + 2
\frac{\gamma+1}{x} + \frac{u'(x)}{u(x)} \right)\;.
\end{equation}

In ref.\ \cite{5} we have shown that the symmetry algebra underlying the
eigenvalue problem associated with $H_-$ is a non-linear one. To be more
explicit, with the help of the ladder operators for $H_+$ given by 
$c= (\rmd/\rmd x+x)^2/2-(\gamma + 1)(\gamma + 2)/2x^2$, 
which together with its adjoint $c^\dagger$ and $H_+$ close a
(linear) Lie algebra, one can introduce similar ladder operators for $H_-$
defined by $D = A^{\dagger} cA$ and its adjoint $D^{\dagger} = A^{\dagger}
c^{\dagger}A$. These operators act on eigenstates of $H_-$ as follows:
\begin{equation}
\fl D^{\dagger} \psi^{-}_{n} (x) = f_{n+1} \psi^{-}_{n+1} (x)\; , \quad
D\psi^{-}_{n} (x)= f_{n} \psi^{-}_{n-1} (x) \; ,\quad D \psi^{-}_{0} (x) = 0\;,
\end{equation}
where $f_{n}$ is given by
\begin{equation}
f_{n} = -2 \sqrt{\textstyle n(n+\gamma+\frac{1}{2})(2n+2\gamma+2+\varepsilon)
(2n+2\gamma+\varepsilon)} \;.
\end{equation}
From these relations it also follows that
\begin{equation}
\fl
\begin{array}{ll}
\psi^{-}_{n}(x)&=\displaystyle \left(f_1 f_2 \cdots f_n\right)^{-1}
\left(D^{\dagger}\right)^n \psi^{-}_{0}(x)\\[2mm]
& =
\textstyle (-\frac{1}{4})^n\left[n!\,(\gamma + \frac{3}{2})_n
(\gamma + 1 + \frac{\varepsilon}{2})_n (\gamma + 2 + \frac{\varepsilon}{2})_n 
\right]^{-1/2} (D^{\dagger})^n \psi^{-}_{0} (x)\;.
\end{array}
\end{equation}
The non-linear algebra closed by the operators $D$, $D^\dagger$ and $H_{-}$
explicitly reads  
\begin{equation}
\left[H_{-},D\right]=-2D \;,\quad 
\left[H_{-},D^{\dagger}\right]= 2D^{\dagger}\;,\quad 
\left[D,D^{\dagger}\right]= \Phi (H_{-})\;,
\end{equation}
where the non-linear structure function\footnote{We note that in the case of
  Lie algebras the structure function would have been a linear one.}
$\Phi$ is given by
\begin{equation}
\textstyle
\Phi (H_{-}) = 8 H_{-}^{3} - 12 (\gamma + \varepsilon + \frac{1}{2}) H_{-}^{2}
+ 4 (2 \varepsilon \gamma + \varepsilon^2 + \varepsilon + 1) H_{-}\;.
\end{equation}
Actually, these types of algebras (having as structure function a polynomial of
degree $p-1$ in one of the generators) are called $W_p$ algebras. More
explicitly, the above algebra (14) is a polynomial deformed $su(1,1)$ algebra
and 
has first been discussed in some detail by Ro\u{c}ek \cite{Roc91}. For a
discussion within a more general approach see also Karassiov \cite{Ka94} and
Katriel and Quesne \cite{KaQu96}.

The quadratic Casimir operator for the non-linear (cubic) algebra (14) reads
\begin{equation}
C = DD^{\dagger} - \Psi (H_{-}) 
\end{equation}
where
\begin{equation}
\Phi (H_{-})=\Psi (H_{-})-\Psi (H_{-}-2)\;.
\end{equation}
We note that in the above Fock space representation (11)-(13) we have the
relations 
\begin{equation}
\fl  \begin{array}{l}
\Psi (H_{-})= f^2_{H_-/2 -\gamma -\varepsilon/2} =
(H_--2\gamma-\varepsilon)(H_-+1+\varepsilon)(H_-+2)H_- \;,\\[2mm]
DD^\dagger = \Psi (H_{-})\;,\quad D^\dagger D= \Psi (H_{-}-2)\;,
  \end{array} 
\end{equation}
and, therefore, the Casimir operator (16) vanishes as expected
\cite{Ka94,KaQu96}. This, however, will in general not be the case 
for non-Fock space representations of the algebra (14)
\cite{Roc91,Ka94,KaQu96}. 

\section{The non-linear coherent states}
We shall now construct coherent states corresponding to the algebra in (14). At
this point we note that coherent states can be constructed following any of the
three methods \cite{ZFG90}: (i) By applying the unitary displacement operator
to the ground 
state. (ii) Defining coherent states as eigenstate of the lowering operator.
(iii) Defining coherent states as minimum uncertainty states. These three
methods are generally not equivalent and only in the case of the standard
harmonic oscillator, where the commutator of the raising and lowering operator
is the unit operator, these three methods are equivalent. Since the symmetry
algebra in the present case is a non-linear one, the Baker-Campbell-Hausdorff
disentangling theorem cannot be used 
and so we shall follow the second approach to construct non-linear coherent
states. Note that coherent states obtained in this way are essentially
Barut-Girardello coherent states \cite{7a}. We also remark that the procedure
following below is very similar to the construction of coherent states
associated with quantum groups \cite{Od98}.

Thus we define coherent states as 
\begin{equation}
| \mu \rangle = \sum^{\infty}_{n=0} c_n \,\mu^{n}\, | n \rangle\;,
\label{19}
\end{equation}
where the $c_n$'s are real constants to be determined, $\mu$ is an arbitrary
complex number, and the ket $| n \rangle$ is a short-hand notation for the  
 eigenstate $\psi^{-}_{n}$ of $H_-$. 
Now, by our definition $|\mu\rangle$ should be an eigenstate of the lowering
operator $D$ and so we have 
\begin{equation}
D|\mu\rangle = \mu\, |\mu\rangle = \sum^{\infty}_{n=0} c_{n+1}\, \mu^{n+1}\,
f_{n+1}\, |n\rangle \;.
\label{20}
\end{equation}
Comparing this result with definition (19) we obtain the recurrence relation 
\begin{equation}
c_{n+1} = \frac{c_n}{f_{n+1}}\; ,\quad n = 0 , 1 , 2 , \ldots\;,
\end{equation}
and consequently the constants $c_n$ for $n\geq 1$ are given by
\begin{equation}
c_n =c_{0}\,\prod^{n}_{i=1} (f_i)^{-1}\;.
\end{equation}
The remaining constant $c_{0}$ is determined via the normalisation of the
coherent states:  
\begin{equation}
  \begin{array}{ll}
\langle \mu | \mu \rangle &\displaystyle
= c^{2}_{0} \left[1+\sum^{\infty}_{n=1}
\left(\prod^{n}_{i=1} f_i^{-2}\right) |\mu|^{2n}\right]\\[4mm]
&\displaystyle = 
c^{2}_{0}\sum^{\infty}_{n=0}
\frac{(|\mu|^2/16)^n}{n!\,
(\gamma + \frac{3}{2})_n(\gamma + 1 + \frac{\varepsilon}{2})_n 
(\gamma + 2 + \frac{\varepsilon}{2})_n}=
1\;.
  \end{array}
\end{equation}
Hence, the normalisation constant $c_0=c_0(\mu)$ can be expressed in terms of a
generalised hypergeometric function 
\begin{equation}
c^{- 2}_{0}(\mu) = \,{} _{0}F_{3}
\left(\gamma + \frac{3}{2} , \gamma + 1 + \frac{\varepsilon}{2} , \gamma +
2 + \frac{\varepsilon}{2};  \frac{| \mu |^2}{16} \right)\;.
\label{c0}
\end{equation}
Similarly we can show that these non-linear coherent states are not orthogonal
for $\mu\neq\nu$,
\begin{equation}
\langle\mu | \nu\rangle =c_{0}(\mu)c_{0} (\nu) \,{} _{0}F_{3}
\left(\gamma + \frac{3}{2} , \gamma + 1 + \frac{\varepsilon}{2} , \gamma +
2 + \frac{\varepsilon}{2};  \frac{\mu^*\nu}{16} \right)\neq 1\;,
\end{equation}
and, therefore, form an over-complete basis in the Hilbert space. 

Another important property, namely, the resolutions of unity can also be
obtained for these non-linear coherent states. Let us  
assume that we have a positive measure $\rho$ on the complex plane such that
\begin{equation}
  \int\limits_{\mathbb C}\rmd\rho(\mu^*,\mu)\,|\mu\rangle\langle\mu| = 1\;.
\label{unity}
\end{equation}
Making the polar decomposition $\mu=\sqrt{x}\,\rme^{\rmi\varphi}$ and the
ansatz 
$\rmd\rho(\mu^*,\mu)=\rmd\varphi \rmd x\,\sigma(x)/2\pi c_0^2(\sqrt{x})$, 
with $\sigma$ being a yet 
unknown density on the positive half-line, the above resolution of unity (26)
reduces to the relations
\begin{equation}
\fl
\int\limits_0^\infty dx\, x^n\sigma(x)= \textstyle
16^n \,n!\,
(\gamma+\frac{3}{2})_n\,
(\gamma+1+\frac{\varepsilon}{2})_n\,
(\gamma+2+\frac{\varepsilon}{2})_n\;,\quad
n=0,1,2,\ldots\;.
\label{moments}
\end{equation}
In other words, $\sigma$ is a probability density on the positive half-line
defined via the moments given above. For technical details on this so-called
Stieltjes moment problem see \cite{Ak65}. Here we note that the integral
(\ref{moments}) may be viewed as Mellin transformation \cite{Erd54} of the
density $\sigma$. In other words, $\sigma$ is given via the inverse Mellin
transformation of its moments. For the above moments this inverse
transformation leads (see ref \cite{Erd54} p 353) to a Meijer G-function
\cite{Erd53} and we explicitly have
\begin{equation}
\sigma(x)=\frac{
G^{40}_{04}\left(\frac{x}{16}\left|0, \gamma +\frac{1}{2},\gamma +
    \frac{\varepsilon}{2},\gamma + 1 +     \frac{\varepsilon}{2} \right.\right)
}{16\, \Gamma(\gamma +\frac{3}{2})\,\Gamma(\gamma +1+\frac{\varepsilon}{2})\,
\Gamma(\gamma + 2 + \frac{\varepsilon}{2}) }\;.
\label{sigma}
\end{equation}
In Figure 1 we plot the radial density $f(x)=\sigma(x)/c_0^2(\sqrt{x})$ for
fixed $\gamma =1$ and various values of $\varepsilon> -2 \gamma -2$. Figure
2 presents the same quantity now, however, with fixed $\varepsilon=1$
and various values of $\gamma \geq 0$.
\begin{figure}
\vspace{100mm}%
\includegraphics{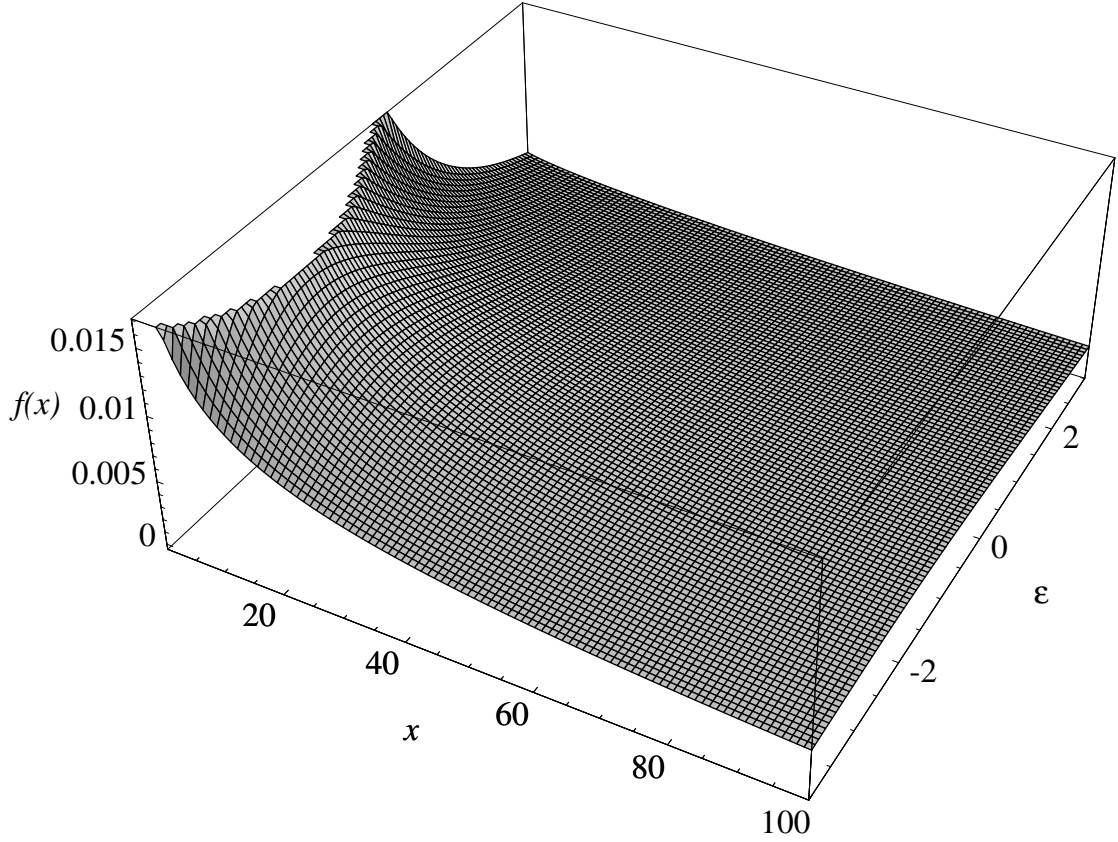}%
\vspace{20mm}%
\caption{A plot of the radial density $f(x)=\sigma(x)/c_0^2(\sqrt{x})$ entering
  the resolution of unity (\ref{unity}) with $\sigma$ and $c_0$ given by
  (\ref{sigma}) and (\ref{c0}), respectively. Here we have fixed the parameter
  $\gamma=1$.}
\end{figure}
\begin{figure}
\vspace{100mm}%
\includegraphics{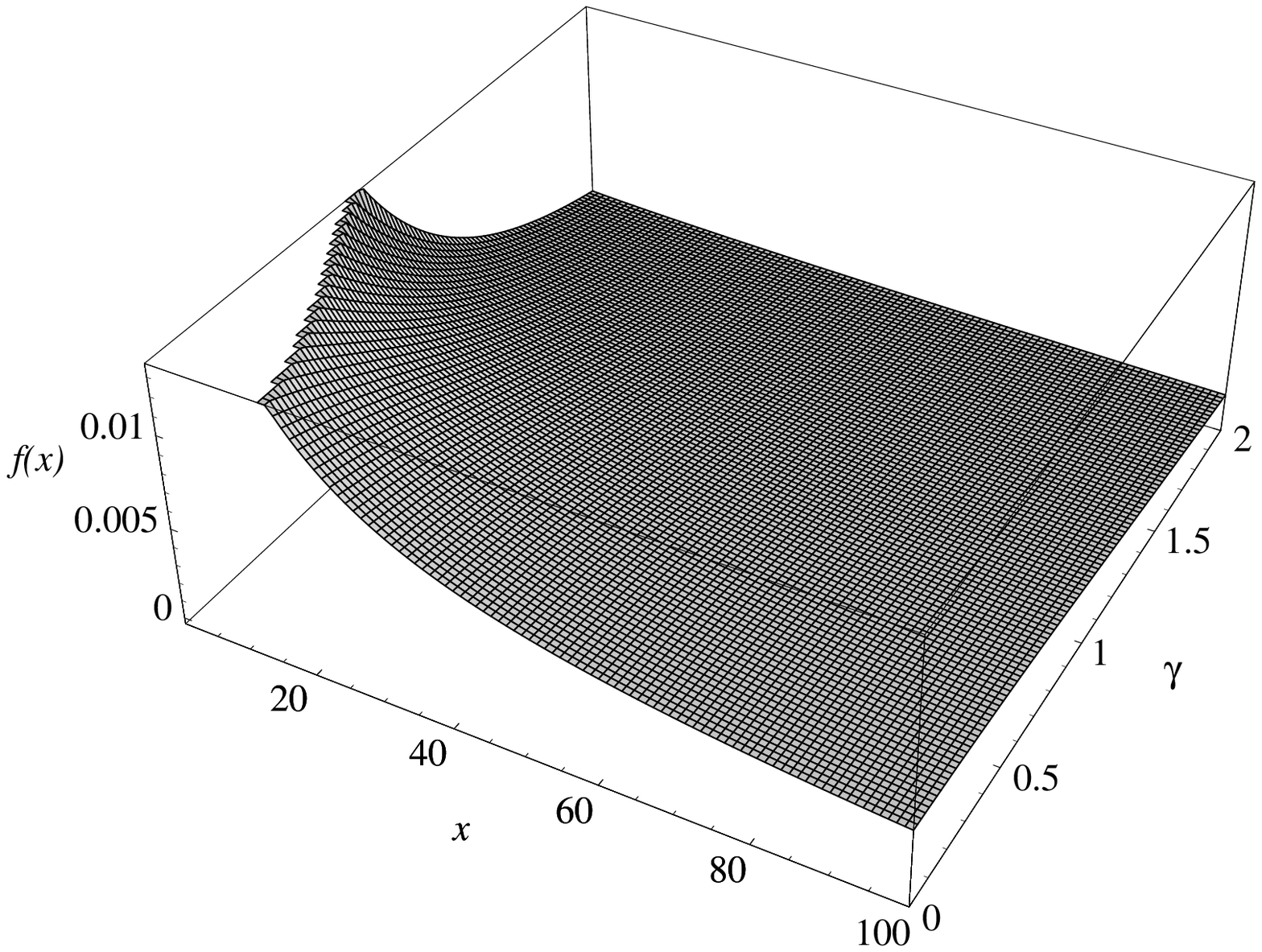}%
\vspace{20mm}%
\caption{Same as Figure 1 but now with fixed $\varepsilon =1$ and various
  values of $\gamma \geq 0$.}
\end{figure}

We now proceed to examine some further properties of these non-linear coherent
states. To do this we define the following hermitian operators:
\begin{equation}
X_1 = \frac{D + D^{\dagger}}{2}\;,\quad X_2 = \frac{D - D^{\dagger}}{2\rmi}\;.
\end{equation}
In terms of these operators the non-linear algebra (14) reads
\begin{equation}
\left[ H_- , X_1 \right] = - 2\rmi X_2 \;,\quad 
\left[ H_- , X_2 \right] = 2\rmi X_1 \;,\quad
\left[ X_1 , X_2 \right] = \frac{\rmi}{2}\, \Phi (H_-) \;.
\label{29}
\end{equation}
The uncertainty relation for the two operators $X_1$ and $X_2$ in some
state $|\psi\rangle\in{\cal H}$ reads
\begin{equation}
(\Delta X_{1})^2_\psi \,(\Delta X_{2})^2_\psi  \geq \frac{1}{4}
\left|\,\langle\psi|[ X_1 , X_2 ]|\psi\rangle\,\right|^2\;,
\label{30}
\end{equation}
where $(\Delta
X_{i})_\psi^2=\langle\psi|X^2_{i}|\psi\rangle-\langle\psi|X_{i}|\psi\rangle^2$.
We note that the non-linear coherent states $|\mu\rangle$ in (\ref{19}) having
property (\ref{20}) always satisfy the equality sign in (\ref{30}).
Note that in the notation used in \cite{12} these states are called
intelligent states.  However, it can be shown that when the functional
$F(\mu)=\left(\langle\mu|DD^{\dagger}|\mu\rangle - |\mu|^2 \right)$ attains
its minimum for some value of $\mu$, say $\mu_0$, then the non-linear coherent
state $|\mu_0\rangle$ is a minimum uncertainty state corresponding to the
non-linear algebra (\ref{29}).

\section{The case of unbroken SUSY}
Let us now briefly describe the situation when SUSY is unbroken. In this case
we choose
\begin{equation}
W(x) = x - \frac{\gamma + 1}{x} +\frac{u'(x)}{u(x)}\;,\quad \gamma \geq 0\;,
\label{newW}
\end{equation}
where now 
$u(x)={}_1F_1(\textstyle\frac{1-\varepsilon}{2},-\gamma-\frac{1}{2},-x^2)$.
For a more general case and the conditions on the parameters $\varepsilon$ and
$\gamma$ see ref \cite{5}. 
It turns out that $V_+$ again represents the radial oscillator while 
$V_-$ is a CES potential. Note that essential details of this problem can
be obtained from the broken SUSY case by replacing $\gamma$ by $- \gamma - 2$.
However, the eigenvalues for $H_-$ are now given by
\begin{equation}
E_{0} = 0\; ,\quad E_{n+1} = 2n + 1 + \varepsilon\;,
\end{equation}
which coincides with the spectrum of $H_+$ with the exception of the
vanishing ground-state energy, which is missing in $H_+$ due to unbroken SUSY.
For the explicit form of the corresponding eigenstates we refer to \cite{5}.

Again we may define ladder operators $D=A^{\dagger}cA$ and 
$D^{\dagger}=A^{\dagger}c^{\dagger}A$ where the SUSY operators $A$ and
$A^{\dagger}$ are now defined with the new SUSY potential (\ref{newW}). They
act on the eigenstates of $H_-$ in the following way: 
\begin{equation}
D^{\dagger}|n+1\rangle=g_{n+1}|n+2\rangle\;,\quad
D|n+1\rangle=g_n|n\rangle\;,\quad
D|0\rangle=0=D^{\dagger}|0\rangle\;,
\label{D}
\end{equation}
where 
\begin{equation}
g_n = -2 \textstyle 
\sqrt{n(n+\gamma+\frac{3}{2})(2n-1+\varepsilon)(2n+1+\varepsilon)}\;.
\end{equation}
From the last relation in (\ref{D}) it is clear that the ground state is
isolated 
in the sense that the non-linear algebra is (non-trivially) realised over the
excited states only. Note that the non-linear algebra closed by $D$,
$D^{\dagger}$ and $H_-$ is identical in form with (14). However, in the
structure function (15) we have to replace $\gamma$ by $-\gamma-2$ \cite{5}.

Now proceeding as in the case of broken SUSY, we can find a superposition state
which is an eigenstate of the annihilation operator $D$. However, this
non-linear coherent state is now given by a superposition of the excited
energy eigenstates:
\begin{equation}
|\eta \rangle= \sum_{n =0}^{\infty} d_n \, \eta^n\, |n + 1\rangle\;,
\label{eta}
\end{equation}
where $\eta$ is a complex number and the $d_n$'s are given by 
\begin{equation}
  \begin{array}{l}
\displaystyle
d_n =d_0\prod^{n}_{i=1} g_i^{-1}\;,\quad n=1,2,3,\ldots\;,\\
d^{-2}_{0}(\eta) = \,{} _{0}F_{3}
\left(\gamma + \frac{5}{2} , \frac{\varepsilon}{2}+\frac{1}{2} ,
  \frac{\varepsilon}{2}+\frac{3}{2};\frac{|\eta |^2}{16} \right) \;.
  \end{array}
\end{equation}
We note that the states $|\eta\rangle$ in (\ref{eta}) are very similar to
coherent 
states although they are not  coherent states. In particular, the states 
$|\eta\rangle$ are not complete because of the absence of the ground state in
the superposition (\ref{eta}). 
We can, however, call these states excited coherent states or photon-added
coherent states \cite{8a} because $|\langle 0| \eta\rangle|^2 = 0$ for all
$\eta\in{\mathbb C}$. Note that $\lim_{\eta\to 0}|\eta\rangle=|1\rangle$. 
The corresponding resolution of unity reads in this case
\begin{equation}
  \int\limits_{\mathbb C}\rmd\rho(\eta^*,\eta)\,|\eta\rangle\langle\eta| =
   1 - |0\rangle\langle 0|\;,
\end{equation}
where $\eta=\sqrt{x}\,\rme^{\rmi\varphi}$, 
$d\rho(\eta^*,\eta)=\rmd\varphi \rmd x\, \sigma(x)/2\pi d_0^2(\sqrt{x})$ 
and the probability density $\sigma$ is again given
via its moments:
\begin{equation}
  \int\limits_0^\infty \rmd x\,x^n \sigma(x)=
\textstyle
16^n \,n!\,
(\gamma+\frac{5}{2})_n\,
(\frac{1}{2}+\frac{\varepsilon}{2})_n\,
(\frac{3}{2}+\frac{\varepsilon}{2})_n\;,\quad
n=0,1,2,\ldots\;.
\end{equation}
As in the case of broken SUSY $\sigma$ can be expressed in terms of a Meijer
G-function and explicitly reads
\begin{equation}
\sigma(x)=\frac{
G^{40}_{04}\left(
\frac{x}{16}\left|0, \gamma +\frac{3}{2},
  \frac{\varepsilon}{2}-\frac{1}{2},\frac{\varepsilon}{2}+\frac{1}{2}
    \right.\right)
}{16\, \Gamma(\gamma +\frac{5}{2})\,\Gamma(\frac{\varepsilon}{2}+\frac{1}{2})\,
\Gamma(\frac{\varepsilon}{2}+\frac{3}{2}) }\;.
\end{equation}

\section{Final remarks}
Starting from the cubic algebra formed by the ladder operators of CES
Hamiltonians related to the radial harmonic oscillator we have constructed the
associated non-linear coherent states. These states are different to those
obtained recently \cite{9} via the Darboux transformation from 
standard (linear) coherent states \cite{6}.
The present non-linear coherent states have been shown to be minimum
uncertainty states with respect to the $X_1$-$X_2$ uncertainty
relation.  

In the present approach we have constructed non-linear coherent states as
eigenstates of the annihilation operator (method (ii)), which turn out to be
equivalent to those defined as minimum uncertainty states (method (iii)).
It would also be of interest to find similar states which equalise other
uncertainties like $H_-$-$X_1$ or $H_-$-$X_2$, and
find their relations to the present one. 
Another interesting possibility is to construct in a similar way coherent
states related to other CES potentials. For example, those related to the CES
potentials which are SUSY partners of the linear harmonic oscillator. Here the
algebra formed by the ladder operators closes a quadratic algebra and SUSY is
unbroken \cite{4,5}. In fact, in doing so \cite{CJT98} one finds other
non-linear coherent states which generalise those previously constructed by
Fern\'andez et al \cite{Fer94}. 


\ack
One of the authors (GJ) would like to thank P.P.\ Kulish, H.\ Leschke, A.\
Odzijewicz, J.\ Trost and S.\ Warzel for valuable comments and suggestions. 


\end{document}